\documentclass[prd,superscriptaddress,onecolumn,showkeys]{revtex4}
\usepackage{eurosym}
\usepackage{amsfonts}
\usepackage{array}
\usepackage{amsthm}
\usepackage{bm}
\usepackage{palatino}
\usepackage{mathpazo}
\usepackage{amssymb}
\usepackage{eurosym}
\usepackage{amsmath}
\usepackage{epsfig}
\usepackage{float}
\usepackage{graphics}
\usepackage{changes}
\usepackage{color}
\usepackage{graphicx}
\usepackage[colorlinks=true,
            linkcolor=blue,
           urlcolor=black,
           citecolor=black]{hyperref}

\def\be{\begin{equation}}
\def\ee{\end{equation}}
\def\beq{\begin{eqnarray}}
\def\eeq{\end{eqnarray}}

\def\bes{\begin{eqnarray}}
\def\ees{\end{eqnarray}}

\newcommand\myeq{\mathrel{\overset{\makebox[0pt]{\mbox{\normalfont\tiny\sffamily $h\rightarrow\infty$}}}{=}}}
\begin{document}

\title{Study of light deflection and shadow from a hairy black hole under the influence of the non-magnetic plasma}

\author{Riasat Ali}
\email{riasatyasin@gmail.com}
\affiliation{Department of Mathematics, Shanghai University and Newtouch Center for Mathematics of Shanghai University, Shanghai-200444, People's Republic of China}

\author{Xia Tiecheng}
\email{xiatc@shu.edu.cn}
\affiliation{Department of Mathematics, Shanghai University and Newtouch Center for Mathematics of Shanghai University, Shanghai-200444, People's Republic of China}

\author{Muhammad Awais}
\email{awaiseducation@gmail.com}
\affiliation{Department of Mathematics, University of Engineering and Technology Lahore-54890, Pakistan}

\author{Rimsha Babar}
\email{rimsha.babar10@gmail.com}
\affiliation{Department of Mathematics, GC
University Faisalabad Layyah Campus, Layyah-31200, Pakistan}

\begin{abstract}
This article computes the bending angle of a hairy black hole at weak field limits. The Gauss-Bonnet theorem is applied to the Gaussian optical curvature; this gives a way to calculate the hairy black hole light bending angle using the Gibbons and Werner approach. We determine the light's deflection angle under the influence of gravitational decoupling parameters, non-magnetic plasma, and dark matter. Further, using the ray tracing phenomenon, we determine the shadow at which light is deflected when a non-magnetic plasma medium is present. There must exist unstable circular light orbits that can act as limit curves for light rays in a spiral motion. In this manner, the shadow can be calculated for an observer at any distance from the center, and the energy emission rate for the hairy black hole can be studied.
\end{abstract}

\date{\today}

\maketitle

\section{Introduction}
Black holes are visible objects typically thought to develop through the gravitational collapse of enormous astronomical objects. The first image of the black hole (BH) in the electromagnetic field in the center of galaxy $M87$ was successfully observed recently \cite{D1, D2}. Since a clear geometrical detection of the BH from the first image is not permitted, examinations will continue to be improved for greater detail in the future \cite{D3}.

Bending light is a well-known classical experiment confirming spacetime geometry's curved nature. It is also well explained in virtually all general relativity (GR) textbooks. The GR predicted that the weak-deflection bending angle \cite{D1a} of starlight brushing the sun is proportional to $(m=\frac{GM}{c^{2}})$, its gravitational radius and the distance of its closest approach 
\begin{equation}
\Psi_{\textit{Einstein}}=4\Big(\frac{m}{r_{0}}\Big)+O\Big(\frac{m}{r_{0}}\Big)^{2}.
\end{equation}
The classic explanation for light deflection is that when a massive body is present, the light ray is deflected by an angle corresponding to its mass within a certain region, known as the impact parameter. A phenomenon called gravitational lensing (GL) occurs when light rays escape and are bent by BHs. The position, angular momentum, and mass of the BHs are only a few of the many details we can learn about them by GL, another invaluable technique. Gravitational lensing has been thoroughly investigated for BHs, cosmic strings, wormholes, and other objects using the geodesic approach ever since the sun's discovery of the light bending angle \cite{D4, D5, D6, D7, D8, D9}. An alternative strategy for obtaining the weak bending angle of light by a spherically symmetric BH in a scenario of optical geometry using the Gauss-Bonnet theorem was presented by Gibbons-Werner in \cite{D10}. Werner examined the Kerr-Randers optical geometry to apply this technique to static BHs \cite{D11}. The weak deflection angle by various BHs in various gravitational theories was extensively explored in Refs. \cite{D12, D13, D14, D15, D16, D17, D18, D19} using Gibbons-Werner's approach. Further, the Gauss-Bonnet theorem (GBT) has been applied recently to study the finite receiver and source of light \cite{D20, D21, D22, D23, D24}. The impact of plasma on light rays can be disregarded in most astronomical scenarios but not in light beams in the microwave range. The impact of the solar corona, which is thought of as a non-magnetic pressure-less plasma, on the angle of deflection and time delay \cite{D25} when light rays propagate close to the sun \cite{D26} is a well-known example. Moreover, many BH solutions in gravity theory and thermodynamic properties (like Hawking temperature and entropy) have been studied in \cite{Q1, Q2, Q3, Q4, Q5}.

Dark matter comprises approximately $27\%$ of the universe's total mass-energy \cite{48}. Dark matter exists to be detected by gravitational effects. It is non-baryonic, non-relativistic, and exhibits weak non-gravitational interactions. Dark matter ideas include WIMPs, super-WIMPs, axions, and sterile neutrinos \cite{49}. The dark-atom framework indicates that dark matter is a composite studied by deflecting light. However ignored, dark matter exhibits electromagnetic properties \cite{50}. The refractive index for frequency-dependent dark matter allows moving photons to detect its optical features. The refractive index implies the speed at which a wave propagates through a medium. Dark matter particles are not electrically charged but can pair with other particles with a hypothetical electromagnetic charge or photons \cite{51, 52, 53, 54}. First, compute the scattering amplitude to determine the amplitude of dark-matter annihilation into two photons. The light refractive index can be obtained, with the actual portion corresponding to propagation speed.

It is widely acknowledged that photons from an illuminating source behind a BH will cause the so-called BH shadow, or the two-dimensional nature of the dark zone, to appear in the observer's sky. The BH shadow, an impression of the BH, informs astronomers of essential details about the BH. Light rays are diverted towards the singularity by the BH gravitational solid pull, which causes those that skim the photon sphere to begin looping around it. A photon will never stop encircling the BH once it lands precisely on the photon sphere. This phenomenon, which affects light rays passing near the unstable photon region, increases the brightness of the first source around the shadow's edge by extending the light ray's path length. As a result, the brightness of the cloud instantly outside the shadow appears to be increased, as well as the image of a dark disc illuminated by a brilliant surround. As a result, it is often described as the crucial curve that shows up as a release ring that is homogeneous and isotropic. The magnitude of the shadow cast by a certain BH is mainly determined by its intrinsic properties; its contour is influenced by light rays and the instabilities of their orbits within the non-magnetic plasma. The first studies of the shadow cast by a perfect circle, or spherically symmetric BH, were conducted by Synge \cite{D26a} and Luminet \cite{D26b}.
Additionally, they presented formulas to determine the shadow's size and angular radius, respectively. Bardeen was the first to study the shadow generated by a Kerr BH \cite{D26c}. The dragging action causes the shadow's shape to be distorted. The BH shadow has since been thoroughly studied in various literary works. For example, Ref. \cite{D26d} studied the BH shadow and photon sphere in dynamically changing spacetimes. Several authors have examined \cite{D27, D28, D29, D30, D31, D32, D33, D34} the unique traces left by alternate gravitational models through the examination of shadows.

The research shows that a BHs bending angle and shadow size are determined by its mass, charge and geometric deformation parameters. This demonstrates the impact of other variables beyond typical Schwarzschild metrics. It combines the effects of a non-magnetic plasma and dark matter on light deflection to provide an advanced perspective of astrophysical phenomena. The work uses the GBT and ray-tracing methods to validate theoretical models under weak-field approximations and broaden their application in astrophysical configurations. The photon sphere's stability and relationship to shadow size are investigated, and results show that photon orbits are unstable near BHs. This is crucial for understanding light behavior near extreme gravitational sources. The results help analyze BH photos acquired by measurements like the Event Horizon Telescope (EHT), including analyzing shadow size and shape. The existence of plasma affects shadow size, resulting in modifications in shadow size estimates depending on plasma density. Dark matter alters light's refractive index, causing dramatic deflection angle changes. This helps refine models for dark matter distribution and its interaction with photons. Insights into light deflection help research of GL, galaxy formation, and the significance of BHs. The laws of physics are broken by classical singularities in GR. Singularities are hidden within a BH's event horizon and can be seen in BH's and the cosmological Big Bang theory. Though singularities are thought to be prohibited by quantum mechanics, the issue of singularities in physics is still unresolved as of this writing. Even within the framework of classical GR, numerous scientists have attempted to create normal BHs to solve the problem of singularities in BHs. This line of analysis discusses the BHs \cite{D35} and Hayward regular BHs \cite{D36}. These BHs were discovered by determining the deflection angles \cite{D37, D38, D39, D40} to investigate the differences with the BH solution in weak bending angle. We introduce parameterized BHs, BH with cosmic strings, and new Schwarzschild BHs and BHs like Einstein Gauss-Bonnet theory into the Keeton-Petters and GBT approaches.

We explored photon deflection as a valuable tool for investigating the physics of BHs. Future astrophysical objects may provide insight into how deformation parameters affect light deflection. The Gaussian curvature of the optical metric from the light beam can be integrated to determine the bending angle using optical geometry. Gibbons-Werner first stated he uses the GBT to determine the BH bending angle in a non-magnetic plasma medium. Optical metrics are constructed to do this, and the GBT is examined. The bending angle in the weak field limit was then achieved by building the optical geometry and applying the GBT. In a non-magnetic plasma medium, the BH bending angle was accurately determined. It is now possible to image supermassive BHs directly due to the EHT. A vital complement to these scientific efforts is general relativistic ray tracing. Images of BHs show a wide variety of physically significant image structures, from immense scales in their relativistic to highly microscopic scales in their photon rings. The BH structure makes it appropriate for a ray-tracing method to create simulated images encompassing the system's essential components. Here, we offer an algorithm for dynamic ray tracing that makes computing highly high-resolution images of BHs more efficient. Any shadow analysis, in addition to GR, would be a fascinating indication.

The structure of this paper is as follows: we start Section {\bf II} by going over some of the fundamental ideas about hairy BH geometry. We determine the bending angle by introducing the optical metric and using the Gaussian optical curvature by the Gibbons-Werner approach in Section {\bf III}. To confirm our findings, we found a hairy BH bending angle in a non-magnetic plasma and dark matter. We study the shadow by applying a ray-tracing approach to the hairy BH and emission energy in Section {\bf IV}. Last, we conclude our findings in Section {\bf V}.

\section{Introductory Review of Hairy Black Holes Solution}

The BH can be fully explained by three externally measured classical aspects like mass, charge, and angular momentum, which is an exciting scenario for Einstein's GR theory \cite{D40a}. The no-hair theory states that a BH should only hold the mass, charge, and angular momentum. Though it was speculated in \cite{D40b} that BHs might have extra charges connected to inner gauge symmetries, it is now established that BHs can have soft quantum hair. The hairy BH solutions could result from the appearance of new primary fields that affect the BH's structure. Black hole solution for hair production instead of focusing on specific essential aspects. In a study of the particular primary fields needed to produce hair in BHs, the research authors \cite{D40c} imagined the existence of a generic source and the source forming the vacuum of the Schwarzschild solution.

The primary goals were to examine a light deflection in the hairy BH context solution and assess how it changed from the standard Schwarzschild context solution. Moreover, the Hairy BH approach has prompted more investigation into its applicability to hairy Kerr \cite{D40d}. The hairy BHs generated by the gravitational decoupling in Ref. \cite{D40c} through the minimal geometric deformation extended are briefly described in this section (see \cite{D42, D43, D44, D45, D46, D47, D48, D49} for further details about minimal geometric deformation and gravitational decoupling). Defined by the hairy BH spacetime \cite{D50} as
\begin{equation}
ds^{2}=F(r)dt^{2}-G(r)dr^{2}-r^{2}\left(d\theta^{2}+\sin^{2}\theta d\phi^{2}\right),\label{a1}
\end{equation}
with
\begin{equation}
F(r)=\frac{1}{G(r)}=1-\frac{2 M}{r}+\frac{Q^{2}}{r^{2}}
-\frac{\alpha (l+M-r)}{r},\nonumber   
\end{equation}
where the geometric deformation of hairy BH parameters are $ (\alpha, Q,l)$ and the BH mass is $M$. It should be noted that $Q$ represents any other charge for the Maxwell tensor or a tidal charge with an extra-dimensional origin. It is not always an electric charge. The solution provides the event horizon by
\begin{equation}
\alpha l=r+\frac{Q^{2}}{r}-2M-M\alpha e^{-r/M}
\end{equation}
To satisfy the dominant energy condition, the $r \geq 2M$, resulting in further constraints on $l$ and $Q$ as
\begin{equation}
Q^{2}\geq \left(\frac{M}{e}\right)^{2}4\alpha,~~~\frac{M}{e^{2}} \leq l.
\end{equation}
The horizon radius can be calculated as
\begin{equation}
r_{\pm}=\frac{\alpha l+2M+\alpha M \pm \sqrt{(\alpha l+2M+\alpha M)^{2}-4(\alpha+1)Q^{2}}}{2(\alpha+1)}\label{RO}
\end{equation}
The Eq. (\ref{RO}) clarifies that the following need must be met for the BH to exist. If
\begin{equation}
(\alpha l+2M+\alpha M)^{2}-4(\alpha+1)Q^{2}\geq 0.
\end{equation}
Using geometric deformation and the dominant energy condition to get a metric solution, It is referred to as charged hairy BH and expands a metric similar to Reissner-Nordstr\" om. It is noticed that the metric (\ref{a1}) is the Schwarzschild BH when $Q=0$ and $\alpha=0$.

\section{Light deflection by the Gibbons-Werner approach}

The charge and geometric deformation parameters of a hairy BH are crucial to comprehending how it deflects light because the electric charge and geometric deformation parameters drastically change the geometry of spacetime surrounding the hairy BH that has a different deflection angle for light rays passing close to it than for an uncharged and without geometric deformation parameters hairy BH. The spacetime geometry directly influences the amount of light bent by the hairy BH gravitational field; it is a crucial factor in the study of GL occurrence. Gibbons-Werner introduces a different, global properties-focused approach to GL theory. In \cite{D50a, D50b}, the light deflection is generally represented schematically by a Gibbon-Werner diagram, which includes the BHs lens, source star, and image. Specifically, we employ the GBT to describe light beams as spatial geodesics of the optical metric and investigate the astrophysically important weak deflection limit outside of the impulse approximation. The light deflection angle for Einstein-Gauss-Bonnet (EGB) gravity in plasma and non-plasma mediums has been investigated in \cite{D51}. It is concluded that GBT analyzed the weak lensing and calculated the photon ray deflection for EGB gravity. In BHs with a magnetic charge parameter, the light deflection in EGB gravity was slightly smaller than that of the uncharged BH. However, a fascinating prediction of light deflection for the future best explains this behavior and further magnetic charge results. Gravitational lensing can provide new insights into developing modified gravity theories and examining BH distances, spins, and other physical properties. The BH geometries allow an examination of the magnetic charge's impact on the light deflection, considering significant indications of physical characteristics. We use GBT to determine a non-plasma medium's charged, hairy BH bending angle. When the source, observer, and null photon are all in the tropical zone, one can conclude that $(\theta =\frac{\pi}{2})$. By inserting $ds^{2}=0$ into metric Eq. (\ref{a1}), the optical metric is obtained as
\begin{equation}
dt^{2}=\frac{G(r)}{F(r)}dr^{2}+\frac{r^{2}}{F(r)}d{\phi}^{2}.\label{a2}
\end{equation}
The above metric can be re-written as follows
\begin{equation}
dt^{2}=\tilde{F}(r)dr^{2}+\tilde{G}(r)d{\phi}^{2},\label{m2}
\end{equation}
where $\tilde{F}(r)=G(r)/F(r)$ and $\tilde{G}(r)=r^2/F(r)$.
By using the optical metric to get the non-zero Christoffel symbols as
\begin{equation*}
\Gamma^{1}_{11}=\frac{\tilde{G}'(r)}{2\tilde{G}(r)}-\frac{\tilde{F}'(r)}{2\tilde{F}(r)},
~~~\Gamma^{2}_{12}=\frac{\tilde{F}(r)}{r}-\frac{\tilde{F}'(r)}{2},~~~
\Gamma^{1}_{22}=\frac{r^{2} \tilde{F}'(r)}{2\tilde{G}(r)\tilde{F}(r)}-\frac{r}{\tilde{G}(r)},
\end{equation*}
with $1$ and $2$ representing the $r$ and $\phi$ coordinates. The optical metric of the Ricci scalar is obtained as
\begin{equation}
R=\frac{2\tilde{F}''(r) \tilde{G}(r) \tilde{F}(r)-\tilde{F}'(r)\tilde{G}'(r)\tilde{F}(r)-\tilde{F}'^{2}(r)\tilde{G}(r)}{2 \tilde{G}^{2}(r) \tilde{F}^{2}(r)}.  \label{a3}
\end{equation}
The Gaussian curvature is defined as
\begin{equation}
\mathbb{K}=\frac{R}{2}.\label{a4}
\end{equation}
For the charge of hairy BH, the Gaussian curvature is obtained by using equation Eq. (\ref{a4}) as
\begin{equation}
\mathbb{K}\simeq -\frac{2 M}{r^{3}}-\frac{  l \alpha}{r^{3}}+\frac{3 M \alpha}{r^{3}}+\frac{3 Q^{2}}{r^{4}}
+\frac{3 l M \alpha}{r^{4}}+\frac{3 Q^{2} \alpha}{r^{4}}
-\frac{6 Q^{2} M}{r^{5}}-\frac{3 l Q^{2} \alpha}{r^{5}}-\frac{3 Q^{2} M \alpha}{r^{5}}+O(l^{2}, Q^{3},M^{2},\alpha^{2}).\label{a5}
\end{equation}
The bending angle for the charged hairy BH solution using GBT may be found in the non-singular domain $N_{e}$ area by utilizing the double integral subscript ${N_{e}}, (\mathbb{K})dS$ as
\begin{equation}
\int\int_{N_{e}}{\mathbb{K} dS}+\oint_{\partial N_{e}}{kdt}+\sum_{i}\varepsilon_{i}=2\pi\varsigma(\mathbb{K}).\label{a6}
\end{equation}
The expression $k=\bar{g}(\nabla_{\dot{\eta}}\dot{\eta},\ddot{\eta})$ represents the geodesic curvature. The single acceleration vector is denoted by $\bar{g}(\dot{\eta},\dot{\eta})=1,~\ddot{\eta}$. The exterior angle at the $ith$ vertex is expressed by $\varepsilon_{i}$ and the associated jump angles drop to $\frac{\pi}{2}$ as $ e\rightarrow\infty$, yielding $\theta_{O}+\theta_{s}\rightarrow\pi$. The Euler signature is the value $\eta(\mathbb{K})=1$. Then, the GBT implies
\begin{equation}
\int\int_{N_{e}}{\mathbb{K}dS}+\oint_{\partial N_{e}}{kdt}+\varepsilon_{i}=2\pi\eta(B_{e}).\label{a7}
\end{equation}
Here, the angle of the total jump is represented by $\varepsilon_{i}=\pi$. We calculate the geodesic curvature for $e\rightarrow\infty$ as
\begin{equation}
k(U_{e})=\mid\nabla\dot{_{U_{e}}}\dot{U_{e}}\mid.\label{a8}
\end{equation}
The geodesic curvature's radial component is expressed as 
\begin{equation}
(\nabla_{\dot{U_{e}}}\dot{U_{e}})^{r}=\dot{U^{\phi}_{e}}\partial_{\phi}\dot{U_{e}^{r}}+\Gamma^{1}_{22}
(\dot{U}^{\phi}_{e})^{2}.\label{a9}
\end{equation}
If $U_{e}:= r(\phi)=e=$ constant for $e$ is large, we get
\begin{equation}
(\nabla_{\dot{U_{e}}}\dot{U_{e}})^{r}\rightarrow\frac{1}{e}.\label{a10}
\end{equation}
Since there is no defect of topology in the geodesic curvature, $k(U_{e})\rightarrow e^{-1}$. It can be represented using the optical metric Eq. (\ref{a2}), $dt = e d\phi$. As a solution, we get
\begin{equation}
k(U_{e})dt=d\phi.\label{a11}
\end{equation}
Using the above expression, we can obtain the equation as
\begin{equation}
\int\int_{N_{e}}{\mathbb{K} dS}+\oint_{\partial N_{e}}{k dt}~~\myeq ~~\int\int_{T\infty}{\mathbb{K} dS}+
\int^{\pi+\psi}_{0}d\phi.\label{a12}
 \end{equation}
The calculation of the $0th$ order light ray in the limit of weak field is $r(t)=\frac{b}{\sin\phi}$. From the equations (\ref{a6}) and (\ref{a12}), we can get the bending angle as
\begin{equation}
\Psi\simeq -\int^{\pi}_{0}\int^{\infty}_{\frac{b}{\sin\phi}}\mathbb{K}\sqrt{det\vec{g}}dr d\phi.\label{13}
\end{equation}
We calculate the $\sqrt{det\vec{g}}$ as
\begin{equation}
\sqrt{det\vec{g}}=r \left(1-\frac{2 M}{r}+\frac{Q^{2}}{r^{2}}
-\frac{\alpha (l_{0}+M-r)}{r}\right)^{\frac{1}{2}}.\label{a14}
\end{equation}
Calculating the angle of light deflection in Eq. (\ref{13}) involves using the Gaussian curvature up to the leading-order terms as
\begin{equation}
\Psi \simeq \frac{4 M}{b}+\frac{2 l \alpha}{b}+\frac{6 M \alpha}{b}-\frac{3 Q^{2}\pi}{4b^{2}}
-\frac{3 l M \alpha\pi}{4b^{2}}-\frac{3 Q^{2} \alpha \pi}{4b^{2}}+\frac{8 Q^{2} M}{3b^{3}}+\frac{4 l Q^{2} \alpha}{3 b^{3}}+\frac{4 Q^{2} M \alpha}{3b^{3}}+O(l^{2},Q^{3},M^{2},\alpha^{2}).
\label{a15}
\end{equation}

The mass of the BH $M$, BH charge $Q$, geometric deformation of hairy BH parameters $\alpha,~l$, and impact parameters all affect the angle of light deflection obtained in a non-plasma medium. We note that the first term in the obtained angle of light deflection is known as the Schwarzschild BH solution \cite{D51}, and the others are caused by the charged nature of the BH and the geometric deformation of hairy parameters. Furthermore, it should be observed that the second term positive sign suggests that this charged BH strong angle of light deflection is greater than that of the Schwarzschild BH, and the term negative sign indicates that this charged BH weak angle of light deflection. 

The effects of $Q$ and $\alpha$ on the angle of light deflection $\Psi$ to impact $b$ are graphically examined.

\begin{figure}[H]
\centering
\includegraphics[width=6cm,height=6cm]{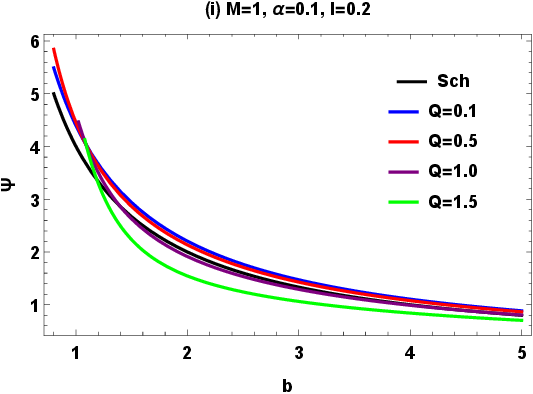}\includegraphics[width=6cm,height=6cm]{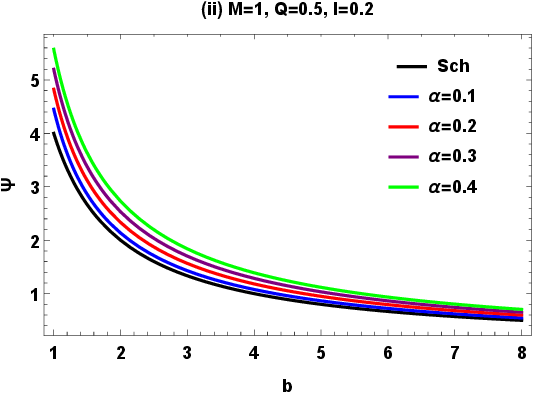}
\caption{angle of light deflection $\Psi$ verse impact parameter $b$ with fixed mass $M=1$ and $l=0.2$ of hairy BH, in the plot of left, using various charge $Q$ and fixed positive $\alpha=0.1$, whereas in the right plot, different $\alpha $ and fixed charge $Q=0.5$.}\label{np}
\end{figure}

Figure {\ref{np} } depicts the angle of light deflection $\Psi$ behavior using the impact parameter $b$ and a fixed BH mass $M=1$ and $l=0.2$. In both plots, one can observe that the angle constantly decreases and then progressively falls within the approximate domain of $0\leq b\leq 5$ in the left plot and $0\leq b\leq 8$ in the right plot. The charge ($Q$) has a minor effect on the angle of light deflection with a fixed deformation parameter ($\alpha$). The deformation parameter ($\alpha$) also has a minor effect on the angle of light deflection with a fixed charge, as seen in both cases of hairy BH. It is evident from both graphs that the deflection angle decreases exponentially with positive values before reaching an asymptotically flat form until it approximates the significant value of $b$. We also observe that the deflection angle is higher than that of the Schwarzschild case in both plots. We also observe that for large values of charge $Q\geq1$, the DA is less than the Schwarzschild case.

\subsection{Light deflection in the non-magnetic plasma}

The deflection angle calculations in the weak-field approximation have restricted uses and should not be applied too near a strongly gravitating object. Therefore, it is preferable to compare findings with specific formulas and establish a range of appropriate radial distances at which it is relevant to apply the weak-field approximation. Several plasma characteristics were determined, and the findings were shown alongside the deflection angle in a vacuum to indicate how much these characteristics appear numerically.
We also compare the ray paths in different plasma models and a vacuum. Thus, the immediate impact of various parameters of a gravitational source on the propagation of rays may be observed.

In this section, we study the angle of light deflection in the effects of non-magnetic plasma. The refractive index for the hairy BH is expressed by
\begin{equation}
n=\sqrt{1-\frac{\omega^{2}_{e}}{\omega^{2}_{p}}F}.\label{a16}
\end{equation}
where $\omega^{2}_{e}$ represents electron plasma frequency and $\omega^{2}_{p}$ represents photon frequency detected at infinity by the observer. For optical metrics, it can be defined as
\begin{equation}
d\sigma^{2}=g^{opt}_{ij}dx^{i}dx^{j}=n^{2}\left(\frac{G}{F}dr^{2}
+\frac{r^{2}}{G}d{\phi}^{2}\right).\label{a17}
\end{equation}
The optical Gaussian curvature in a plasma frame can be calculated by using the formula in Eq. (\ref{a4}) as
\begin{eqnarray}
\mathbb{K} &\simeq& - \frac{2M}{r^3}+\frac{3Q^2}{r^4} - \frac{6M Q^2}{r^5} 
+ \frac{5\omega_e^2 Q^2}{\omega_{p}^2 r^4} 
- \frac{3 \omega_e^2 M}{\omega_{p}^2 r^3} 
- \frac{26 \omega_e^2 M Q^2}{\omega_{p}^2 r^5} 
+ \frac{3 Q^2\alpha}{r^4} - \frac{3 M\alpha}{r^3} - \frac{3 M Q^2\alpha}{r^5} \nonumber
\\&+& \frac{10 \omega_e^2 Q^2\alpha}{\omega_{p}^2 r^4} 
- \frac{15 \omega_e^2 M\alpha}{2 \omega_{p}^2 r^3} 
- \frac{39 \omega_e^2 M Q^2\alpha}{\omega_{p}^2 r^5} 
- \frac{l\alpha}{r^3} - \frac{3 l Q^2\alpha}{r^5} 
+ \frac{3 l M\alpha}{r^4} - \frac{3 l \omega_e^2\alpha}{2 \omega_{p}^2 r^3} 
- \frac{13 l \omega_e^2 Q^2\alpha}{\omega_{p}^2 r^5}\nonumber
\\&+& \frac{12 l \omega_e^2 M\alpha}{\omega_{p}^2 r^4} 
+ \frac{32 l \omega_e^2 M Q^2\alpha}{\omega_{p}^2 r^6}+O(l^{2},Q^{3},M^{2},\alpha^{2}).\label{a18}
\end{eqnarray}
Applying the GBT, we compute the angle of light deflection. To do this, we see the straight line approximation $r(t)=\frac{b}{\sin\phi}$ at zeroth order, and we get
\begin{equation}
\psi_{1}\simeq -\lim_{R\rightarrow 0}\int^{\pi}_{0}\int^{R}_{b/\sin\phi}\mathbb{K}\sqrt{det\vec{g}}dr d\phi.\label{a19}
\end{equation}
For the leading order terms, the hairy BH angle of light deflection under the influence of a non-magnetic medium, we have
\begin{eqnarray}
\Psi &\simeq & \frac{4 M}{b}+\frac{2 l \alpha}{b}+\frac{6 M \alpha}{b}-\frac{3 Q^{2}\pi}{4b^{2}}
-\frac{3 l M \alpha\pi}{4b^{2}}-\frac{3 Q^{2} \alpha \pi}{4b^{2}}
+\frac{8 Q^{2} M}{3b^{3}}+\frac{4 l Q^{2} \alpha}{3 b^{3}}\nonumber\\ &+& \frac{4 Q^{2} M \alpha}{3 b^{3}}+
\frac{6 M \omega^{2}_{e}}{b\omega^{2}_{p}}
+\frac{3 l \alpha \omega^{2}_{e}}{b\omega^{2}_{p}}
+\frac{15 M \alpha \omega^{2}_{e}}{b\omega^{2}_{p}}
-\frac{5 Q^{2}\pi \omega^{2}_{e}}{4b^{2}\omega^{2}_{p}}
-\frac{3 l M \alpha\pi \omega^{2}_{e}}{b^{2}\omega^{2}_{p}}
-\frac{5 Q^{2} \alpha \pi \omega^{2}_{e}}{2b^{2}\omega^{2}_{p}}
\nonumber\\ &+&\frac{104 Q^{2} M \omega^{2}_{e}}{9 b^{3}\omega^{2}_{p}}
+\frac{52 l Q^{2} \alpha \omega^{2}_{e}}{9 b^{3}\omega^{2}_{p}}
+\frac{52 Q^{2} M \alpha \omega^{2}_{e}}{3 b^{3}\omega^{2}_{p}}
-\frac{3  l Q^{2} M \alpha \pi \omega^{2}_{e}}{b^{4}\omega^{2}_{p}}
+O(l^{2},Q^{3},M^{2},\alpha^{2}).\label{a20}
\end{eqnarray}
The angle of light deflection depends on the impact parameter $b$, the geometry of BH $\alpha,~l$, charge $Q$, electron plasma frequency $\omega_e$, and photon frequency $\omega_{p}$ seen to infinity by the observer. If $\frac{\omega^{2}_{e}}{\omega^{2}_{p}} \rightarrow 0$, it can be reduced to Eq. (\ref{a15}) in the non-plasma case.
\begin{figure}[H]
\centering
\includegraphics[width=6cm,height=6cm]{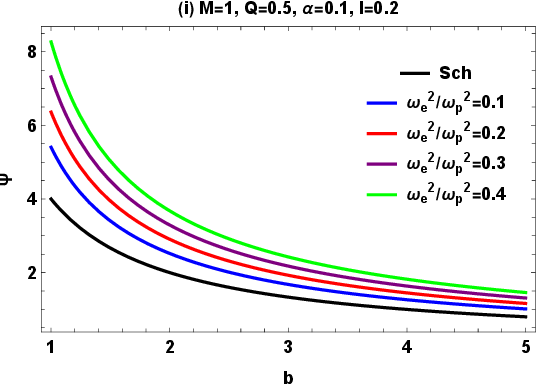}\includegraphics[width=6cm,height=6cm]{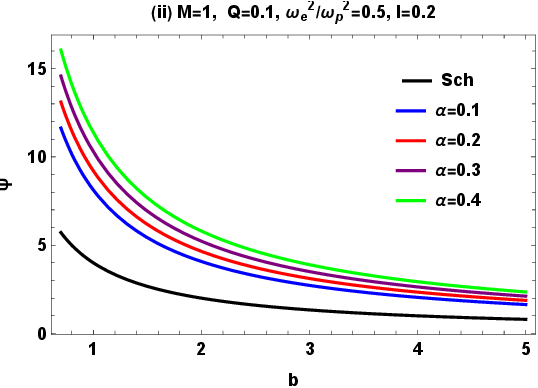}
\caption{the angle of light deflection $\Psi$ verse impact parameter $b$ with fixed BH mass $M=1$, charge $Q=0.5$ and $l=0.2$ is shown. Different plasma frequencies $0.1\leq\omega^2_{e}/\omega^2_{p}\geq 0.4$ and fixed hairy parameters $\alpha=0.1$ are shown on the left, while different deformation parameters $0.1\leq\alpha\geq 0.4$ and fixed $\omega^2_{e}/\omega^2_{p}= 0.5$ are plotted on the right.}\label{P1}
\end{figure}

The behavior of the angle of light deflection $\Psi$ for the impact parameter $b$ is presented in Fig. \ref{P1}, with the different plasma frequencies and different deformation parameters with the BH mass fixed at $M=1$, charge $Q=0.5$ and $l=0.2$. It is evident from both graphs that the angle of light deflection reduces exponentially with positive values before reaching an asymptotically flat form until $b \rightarrow \infty$. It is also possible to see that the angle of light deflection grows with increasing deformation parameters ($\alpha$) and plasma frequency values.
The light deflection angle in the Schwarzschild BH case is lower than that of hairy BH.

\subsection{DEFLECTION ANGLE IN DARK MATTER}

The impact of dark matter on light deflection, which modifies the refraction index due to the presence of dark matter that is a characteristic product of the dark matter distribution, is another feature that may be used to verify the model with our results in Eq. (\ref{a15}). It would be fascinating to compare the simple case deflection angle in scalar hairy BH expected for our dark matter distribution. Here, however, by using the conventional formalism of the refraction index with the dark matter-like field, we can offer some considerations regarding the impact of dark matter on the amount of light deflection. The superposition of the deflection angles of several tiny point masses characterizes lensing in \cite{74}. We can thus obtain a description of light rays in a gravitational field similar to that of classical optics by applying Fermat's principle to the geodesic trajectories of our four-dimensional curved spacetime. Specifically, when Fermat's principle holds, the optical length is minimized for a transparent medium with a constant refractive index. A scalar hairy BH encircled by dark matter could create the light deflection as a scalar hair in a vacuum and show that the presence of dark matter significantly modifies the image of scalar hairy BH, indicating that the mass estimation of distant supermassive scalar hairy BH may vary if they are submerged in dark matter. We suggest exploring the refractive index associated with dark matter lensing in more realistic scenarios. Lastly, we demonstrate how the findings could contribute to understanding supermassive scalar hairy BH brightness. The interaction between the scalar hairy BH and the dark matter in its immediate surroundings is crucial to understanding astrophysical phenomena, including radiation emission. Many observable features can be significantly affected by the presence of dark matter, including the electromagnetic spectrum and scalar hairy BH absorption processes. We employ realistic mathematical models that accurately represent the light deflection surrounding dark matter. The theory is refined by analyzing the interplay between dark matter and hairy BH. It illuminates how various components of the hairy BH system interact and impact each other.

To investigate weak deflection through dark matter, we consider the propagation effects when low mass and greater number density dark matter particles such as warm thermal particles or axion-like particles have relative to ordinary matter. If dark matter interacts with photons purely through quantum fluctuations, the consequence is a refractive index. At very low photon energies ($\mathcal{M}_{f}\sim e^{2}\epsilon^{2}$), the refractive index along with Compton amplitude are correlated \cite{75} as

\begin{equation}
n=1+\frac{\zeta}{4m^{2}_{dm}\omega^{2}_{p}}\mathcal{M}_{f},
\end{equation}
where the dark matter density is $\zeta=1.1\times 10^{-6}GeV/cm^{3}$, and the observed photon frequency is $\omega_{p}$ \cite{75}. If we disregard spin, the amplitude for photon energies below the flexible limit is a definite and even function of $\omega_{p}$. On the other hand, the $O(\omega^{2n}_{p})$ parts have coefficients that are positive, and odd powers in the expansion rate at $\omega_{p}$ may arise from motion-dependent correlations. Consequently, the refractive index transforms to
\begin{equation}
n=1+\frac{\zeta}{4m^{2}_{dm}}\left[\frac{\alpha_{0}}{\omega^{2}_{p}}+\alpha_{1}+\alpha_{2}\omega^{2}_{p}+O(\omega^{4}_{p})\right].
\end{equation}
Dispersive mechanisms in dark matter can bend the photons and also select the refractive index $n(\omega_{p})$ that was corrected by the scattered amplitude of dark matter and light matter \cite{75}. We should use the scalar hairy BH, the refractive index to perform these calculations. In dark matter medium, the scalar hairy BH is expressed \cite{76} as
\begin{equation}
\breve{n}=1+c \alpha_{0}+\alpha_{2}\omega^{2}_{p},\label{a22}
\end{equation}
where $\zeta$ denotes the dispersive particles mass density in dark matter, $\alpha_2\geq0$, $\alpha_0=-2e^2 \epsilon^2$ and $c=\frac{\zeta}{4 m^2_{dm} \omega_p^2}$.
We determine the Gaussian curvature $\mathbb{K}$ at the dark matter by combining the refractive index and the dimensions optical geometry of scalar hairy BH as
\begin{align}
\mathbb{K} &\simeq  - \frac{2M}{r^3(\alpha_{2}\omega^2_p+c \alpha_{0}+1)^{2}}
-\frac{3 M\alpha}{r^3(\alpha_{2}\omega^2_p+c \alpha_{0}+1)^{2}}
- \frac{l\alpha}{r^3(\alpha_{2}\omega^2_p+c \alpha_{0}+1)^{2}}
+ \frac{3 l M\alpha}{r^4(\alpha_{2}\omega^2_p+c \alpha_{0}+1)^{2}}\nonumber 
\\&\quad+\frac{3Q^2}{r^4(\alpha_{2}\omega^2_p+c \alpha_{0}+1)^{2}} 
+ \frac{3 Q^2\alpha}{r^4(\alpha_{2}\omega^2_p+c \alpha_{0}+1)^{2}}
- \frac{3 M Q^2\alpha}{r^5(\alpha_{2}\omega^2_p+c \alpha_{0}+1)^{2}} 
- \frac{3 l Q^2\alpha}{r^5(\alpha_{2}\omega^2_p+c \alpha_{0}+1)^{2}}\nonumber
\\&\quad-\frac{6M Q^2}{r^5(\alpha_{2}\omega^2_p+c \alpha_{0}+1)^{2}}+O(Q^{4}, M^{2},l^{2}, \alpha^2).\label{q1}
\end{align}
When dark matter is present, the angle $\Psi$ can be derived as
\begin{eqnarray}
\Psi &\simeq&  \frac{4 M}{b(\alpha_{2}\omega^2_p+c \alpha_{0}+1)^{2}}
+\frac{2 l \alpha}{b(\alpha_{2}\omega^2_p+c \alpha_{0}+1)^{2}}
+\frac{6 M \alpha}{b(\alpha_{2}\omega^2_p+c \alpha_{0}+1)^{2}} 
-\frac{3 Q^{2}\pi}{4b^{2}(\alpha_{2}\omega^2_p+c \alpha_{0}+1)^{2}}\nonumber \\
&-&\frac{3 l M \alpha\pi}{4b^{2}(\alpha_{2}\omega^2_p+c \alpha_{0}+1)^{2}}
-\frac{3 Q^{2} \alpha \pi}{4b^{2}(\alpha_{2}\omega^2_p+c \alpha_{0}+1)^{2}} 
+\frac{8 Q^{2} M}{3b^{3}(\alpha_{2}\omega^2_p
+c \alpha_{0}+1)^{2}}
+\frac{4 l Q^{2} \alpha}{3 b^{3}(\alpha_{2}\omega^2_p+c \alpha_{0}+1)^{2}}\nonumber \\
&+&\frac{4 Q^{2} M \alpha}{3 b^{3}(\alpha_{2}\omega^2_p+c \alpha_{0}+1)^{2}} + O(Q^{4}, M^{2},l^{2}, \alpha^2). \label{q2}
\end{eqnarray}
When $(\alpha_{2}\omega^{2}_{p}+c \alpha_{0}+1)^{2}$ is substituted to one, then the equation comparable to Eq. (\ref{a15}). Furthermore, the first term corresponds with Schwardchild BH for $Q=0$ and $\alpha=0$. Our calculation may be applied to celestial bodies. For instance, if our goal is to accurately match our analytical formulations to the observed dark matter light deflection effect. This study is ideally only used for astronomical systems in which dark matter is considered rather than other common factors that affect the mass parameter. As a result, we believe that evaluations of the light deflection effects of BHs and regular stars should be good places to apply the findings of this study. However, because dark matter dominates, observations of galaxies or their clusters should be considered less significant. The light deflection of BHs surrounded by perfectly fluid dark matter in literature has been analyzed in \cite{77, 78, 79}.

\begin{figure}[H]
\centering
\includegraphics[width=6cm,height=6cm]{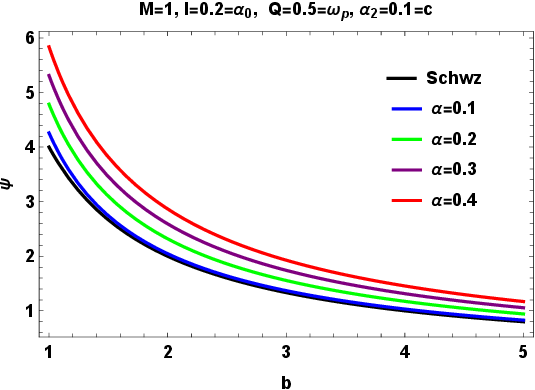}\includegraphics[width=6cm,height=6cm]{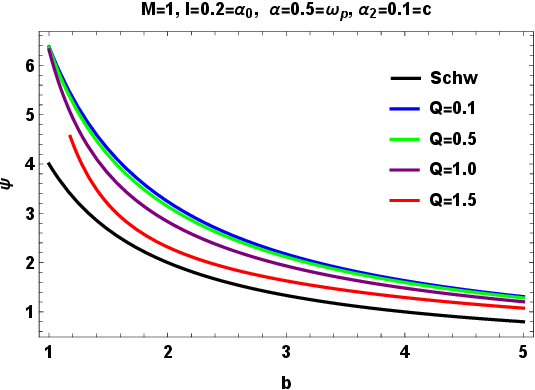}
\caption{the angle of light deflection $\Psi$ in dark medium verse impact parameter $b$ with fixed BH mass $M=1$, $l=0.2=\alpha_0$, $\omega_p=0.5$ and $\alpha_2=0.1=c$ for variations of BH geometry parameter $\alpha$ and charge $Q$.}\label{d1}
\end{figure}

Fig. \ref{d1} shows, for constant values of mass $M=1$, arbitrary parameters $\alpha_2=0.1=c,~\alpha_0=0.2$, photon frequency $\omega_p=0.1$, charge $Q=0.5$, and BH geometry parameter $l=0.2$, the effects of parameter $\alpha$ on deflection angle $\psi$. However, with a constant mass $M=1$, arbitrary parameters $\alpha_2=0.1=c,~\alpha_0=0.2$, photon frequency $\omega_p=0.1$, $Q=0.5$, BH geometry parameters $l=0.2$ and $\alpha=0.5$, the effects of the charge $Q$ on the deflection angle are seen in the right panel. The left panel of the picture illustrates how the deflection angle increases with an increase in parameter $\alpha$ and decreases exponentially with an increase in impact parameter $b$. The deflection angle decreases as the impact parameter increases, and this relationship is inversely related to the charge $Q$ increasing in the right figure.

Both plots exhibit strong GL at low impact parameter $b$ values. Furthermore, we note that weak and low deflection in the dark media is equivalent to that of the non-plasma and plasma mediums. The hairy BH instance has a greater deflection angle than the Schwarzschild BH example.

\section{Shadow in the non-magnetic plasma}

The shadow may be calculated geometrically analytically, at least for spacetimes where the equation for light, such as geodesics, is fully integrable. Almost all of the previous work \cite{RA4} that considered the influence of matter was based on numerical calculations. This work aims to investigate shadows in matter using analytical computations. While this is only possible in idealized conditions, we believe it is still valid. Analytical results provide a more detailed understanding of how an impact is influenced by specific parameters compared to numerical simulations that merely show a particular scenario. Analytical results are valuable for testing numerical codes. This research analyzes how a non-magnetic pressure-less plasma affects the size of a BH shadow. In other words, depending on the frequency, light rays in a plasma deviate from light like geodesics in a dispersive medium. From Maxwell's equations, the light-ray Hamiltonian may be obtained. Two charged fluids generate the electromagnetic field, one representing ions, and the other electrons. A two-scale approach is required to convert Maxwell's equations for a plasma on a curved background to ray optics. Breuer and Ehlers \cite{D52, D53} considered a magnetic pressure-less plasma and gave a formal derivation of the Hamiltonian for light rays. A comparable derivation for the considerably simpler situation of a non-magnetic pressureless plasma can be found in Perlick \cite{D54}. In the latter scenario, a scalar index of refraction, which is a direction-independent index that depends only on the spacetime point and frequency, can adequately describe the equation of light rays. The light-ray equation of motion that results then falls within a class covered in Synge's textbook \cite{D55}. In the Schwarzschild and Ellis wormhole solutions, the shadow is broadly studied in the ray-tracing approach, whose shadow is a function of the radius coordinate \cite{D56}.

The null geodesic equations for the Kerr BH in a vacuum are entirely separable, yielding four constants of motion and a wonderfully simplified issue \cite{D57}. However, the Carter constant is typically absent, and the Hamilton-Jacobi equations for the light rays cannot be separated for generic non-Kerr rotating BHs or Kerr BHs encircled by ionized matter. It is, therefore, challenging to compute analytically. Interestingly, Perlick and Tsupko discovered that to ensure the presence of a generalized Carter constant, the plasma distribution must meet a particular condition \cite{D58}. In \cite{D58}, the effect of plasma on the shadow of a Kerr BH was studied numerically using the photon's motion around Kerr BHs in the presence of radially dense plasma by a ray-tracing approach.

We examine ray tracing numerically, considering a medium's impact on light propagation. A specific type of medium exists a non-magnetic and pressure-less electron-ion plasma, whose impact on the shadow can be addressed analytically and examined logically for hairy spacetime. We present the findings from the computation of the medium influence on the shadow. We operate within the geometry framework to model light propagation regarding rays. This method only makes the plasma noticeable when it affects the ray's hairy spacetime trajectories. The study of shadow from a new type of hairy BH by deriving the equations of motion and taking into account the effect of an electron plasma frequency $\omega_{e}(r)$ on a non-magnetized cold plasma through the Hamiltonian \cite{D59, RA2, RA3} of the moving photon to obtain the conditions as
\begin{equation}
H=\frac{1}{2}\left(g^{ik}j_{i}j_{k}+\omega^{2}_{p}(r)\right)=
\frac{1}{2}\left(-\frac{v^{2}_{t}}{F(r)}
+\frac{v^{2}_{r}}{G(r)}+\frac{v^{2}_{\phi}}{r^{2}}
+\omega^{2}_{p}(r)\right).\label{a22}
\end{equation}
The equations of motion for photons can be obtained by deriving the Hamiltonian Eq. (\ref{a22}) for a two-fluid source from Maxwell's equations. The continuous motion quantities $-v_{t} = E$ and $v_{\phi}$, which are associated with the energy $-v_{t} = E$ and angular momentum $v_{\phi} = L$, can be described as
\begin{equation}
\dot{v_{i}}=-\frac{\partial H}{\partial x^{i}}, ~~~\dot{ x^{i}}=-\frac{\partial H}{\partial v_{i}}\label{a23},
\end{equation}
implies that
\begin{eqnarray}
\dot{v_{t}}&=&-\frac{\partial H}{\partial t}=0,\label{a24}\\
\dot{v_{\phi}}&=&-\frac{\partial H}{\partial \phi}=0,\label{a25}\\ 
\dot{v_{r}}&=&-\frac{\partial H}{\partial r}=0,\label{a26}
\end{eqnarray}
where $v_{r}$ represents the radial momentum and is given by
\begin{equation}
\dot{v_{r}}=-\frac{\partial H}{\partial r}=\frac{1}{2}\left(-\frac{v^{2}_{t}F'(r)}{F^{2}(r)}
+\frac{v^{2}_{r}G'(r)}{G^{2}(r)}
-\frac{d}{dr}\frac{v_{\phi}^{2}}{r^{2}}
-\frac{d}{dr}\omega^{2}_{p}(r)\right),\label{a27}
\end{equation}
and
\begin{eqnarray}
\dot{t}&=&\frac{\partial H}{\partial v_{t}}=-\frac{v_{t}}{F(r)}, \label{a28}\\
\dot{\phi}&=&\frac{\partial H}{\partial v_{\phi}}=\frac{v_{\phi}}{r^{2}}, \label{a29}\\
\dot{r}&=&\frac{\partial H}{\partial v_{r}}=\frac{v_{r}}{G(r)},\label{a30}
\end{eqnarray}
putting $H=0$ then
\begin{equation}
0=-\frac{v^{2}_{t}}{F(r)}
+\frac{v^{2}_{r}}{G(r)}+\frac{v^{2}_{\phi}}{r^{2}}
+\omega^{2}_{p}(r).\label{a31}
\end{equation}
In this case, a prime denotes differentiation about $r$, while a dot denotes differentiation concerning an affine parameter $\lambda$.

It may be inferred from a system of Eqs. (\ref{a24}) and (\ref{a25}) that $v_{t}$ and $v_{\phi}$ are motion constants and $\omega_{0}=-v_{t}$ is written. The formula converts the frequency $\omega(r)$ observed by a static observer into a function of $r$. We have
\begin{equation}
\omega(r)=\frac{\omega_{0}}{\sqrt{F(r)}}.\label{a32}
\end{equation}
We utilize Eqs. (\ref{a29}) and (\ref{a30}) to determine the orbit equation as
\begin{equation}
\frac{dr}{d\phi}=\frac{\dot{r}}{\dot{\phi}}
=\frac{r^{2}v_{r}}{G(r)v_{\phi}}.\label{a33}
\end{equation}
By substituting for $v_{r}$ from Eq. (\ref{a31}), we get
\begin{equation}
\frac{dr}{d\phi}=\pm\frac{r}{\sqrt{G(r)}}
\sqrt{\frac{ \omega^{2}_{0}S(r)^{2}}{v^{2}_{\phi}}-1},\label{a34}
\end{equation}
With the frequency $\omega_{0}$ measured by a static
observer, we have defined the function
\begin{equation}
S^{2}(r)=\frac{r^{2}}{F(r)}
\left(1-F(r)\frac{\omega_{p}^{2}}{\omega_{0}^{2}}\right).\label{a35}
\end{equation}
Given that $X$ denotes the trajectory's turning point, $\frac{dr}{d\phi}|_{X} = 0$ must hold. This formula associates $X$ with the motion constant $\frac{v_{\phi}}{\omega_{0}}$ as
\begin{equation}
S^{2}(X)=\frac{v_{\phi}^{2}}{\omega_{0}^{2}}.\label{a36}
\end{equation}
A stationary observer at point $r_{0}$ is expected to shoot light beams into the past with an angular radius $X$ concerning the radial direction as
\begin{equation}
\cot\beta=
\pm\frac{\sqrt{G(r)}}{r}\frac{dr}{d\phi}|_{r=r_{0}}.\label{a37}
\end{equation}
The orbit Eq. (\ref{a34}) can be rewritten as follows using Eq. (\ref{a36}) in the case that the light beam vanishes after approaching a minimal radius $X$ as
\begin{equation}
\frac{dr}{d\phi}=
\pm\frac{r}{\sqrt{G(r)}}
\sqrt{\frac{S^{2}(r)}{S^{2}(X)}-1}.\label{a38}
\end{equation}
For the angle $\beta$, we obtain
\begin{equation}
\cot^{2}\beta=
\frac{S^{2}(r_{0})}{S^{2}(X)}-1,\label{a39}
\end{equation}
and implies that
\begin{equation}
\sin^{2}\beta=\frac{S^{2}(X)}{S^{2}(r_{0})}.\label{a41}
\end{equation}
The shadow $\beta$ is defined by rays that asymptotically approach a circular light orbit with radius $r_{ph}$. The angular radius of the shadow is obtained as follows from $X \rightarrow r_{ph}$ in Eq. (\ref{a41}); we get
\begin{equation}
\sin^{2}\beta=\frac{S^{2}(r_{ph})}{S^{2}(r_{0})},\label{a42}
\end{equation}
where the formula (\ref{a35}) yields $v(r)$. In many cases, it is conceivable to assume that the observer is situated in an area with a negligibly modest plasma density. Thus, the Eq. (\ref{a35}) yields
\begin{equation}
S^{2}(r)=\frac{r^{2}_{0}}{F(r_{0})},\label{a42}
\end{equation}
From the Eq. (\ref{a41}), we get
\begin{eqnarray}
\sin^{2}\beta &=&
\frac{r^{2}_{ph}(Q^{2}-r_{0}(2M+\alpha(l+M-r_{0})-r_{0})) (r_{ph}(2M+\alpha(1+M))\omega^{2}_{p}-Q^{2}\omega^{2}_{p}-r^{2}_{ph}((1+\alpha)\omega^{2}_{p}-\omega^{2}_{0}))}{r^{2}_{ph}(1+\alpha)-r_{0}(2M+\alpha(1+M)+Q^{2})r^{4}_{0}\omega^{2}_{0}}.\label{a44}
\end{eqnarray}
The mass of the BH, BH charge, geometric deformation of hairy BH parameters, frequency of static observer, and circular light orbit radius all affect the shadows obtained in a non-magnetic plasma medium. We note that the $Q=0$, $\omega_{p}=0$, $\alpha=0$ and $r_{ph}=3M$ in the obtained shadow is known as the Schwarzschild BH \cite{D56}. The hairy BH is usually encircled by non-magnetic plasma, which modifies the shadows' apparent size. The impact of geometrical deformations on BH shadows and photon sphere radius has become evident in hairy spacetimes. Using a ray tracing approach, the shadows from Ellis wormhole space-times \cite{D56} have been computed. We have demonstrated that its impact on the rays, hairy spacetime trajectories, and non-magnetic plasma on BH geometry is insignificant. By comparing our findings with those of Schwarzschild spacetime, which accurately defines the impact of non-magnetic plasma on the apparent form of the BH image, this approach offers a significant understanding of these dynamic astrophysical phenomena.

\begin{figure}[H]
\centering
\includegraphics[width=6cm,height=6cm]{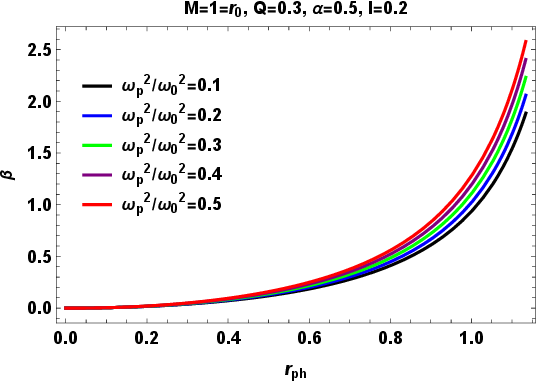}\includegraphics[width=6cm,height=6cm]{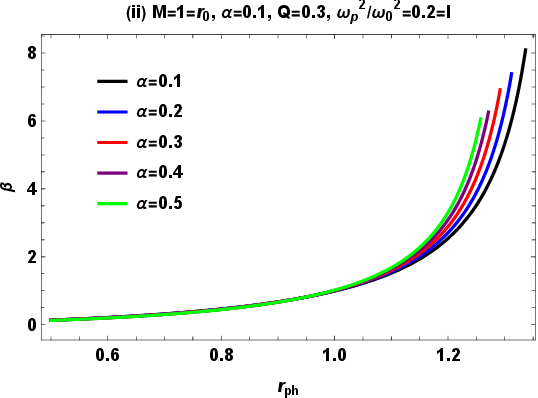}
\caption{The shadow of a hairy BH concerning photon radius $r_{ph}$ with fixed BH mass $M=1$, observer radius $r_0=1$, charge $Q=0.3$ and $l=0.2$ is shown. Different plasma frequencies $0.1\leq\omega^2_{p}/\omega^2_{0}\geq 0.5$ are shown on the left, while different deformation parameters $0.1\leq\alpha\geq 0.5$ are plotted on the right.}\label{S1}
\end{figure}
The changes in shadow $\beta$ for the photon $r_{ph}$ for various selections of the ratio between plasma frequencies and deformation parameters are depicted in Fig. {\ref{S1}}. 
In (i), if $M=1=r_0,~Q=0.3,~\alpha=0.5$, and $l=0.2$ are constant, the shadow rises exponentially for various changes of the plasma frequencies ratio $\omega^2_{p}/\omega^2_{0}$ in the area $0\leq r_{ph}\leq 1.2$. The shadow becomes notably more significant with higher plasma frequency ratio values. The propagation in shadow $\beta$ as $r_{ph}$ grows may be explained by this direct relation between shadow and photon radius.

In the area $0.5\leq r_{0}\leq 1.3$, (ii) shows how shadow $\beta$ behaves in terms of the photon radius $r_{ph}$ for various variations of $\alpha$ with fixed $M=1=r_0,~\alpha=0.1,~Q=0.3$ and $\omega^2_{p}/\omega^2_{0}=0.2=l$. The shadow continuously grows for various values of $\alpha$, resulting in a propagation in the BH shadow radius. The shadow is notably bigger for larger values of the deformation parameter $\alpha$.

When plasma's impact decreases $(\omega^2_e/\omega^2_p\to 0)$, the shadow size approximates that of a Schwarzschild BH, defined only by mass and geometric deformation parameters. The shadow radius is proportional to the photon radius $(r_{ph})$, which is determined by the plasma frequency and BH shape. Higher deformation parameters ($\alpha$) result in an enormous shadow, whereas lower values cause a decrease in size.

The contour plots for the shadow of hairy BH are analyzed in celestial coordinates $x$ and $y$.
\begin{figure}[H]
\centering
\includegraphics[width=6cm,height=6cm]{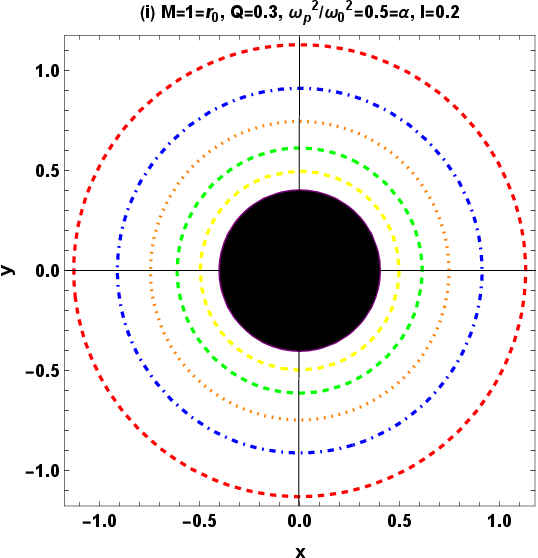}\includegraphics[width=6cm,height=6cm]{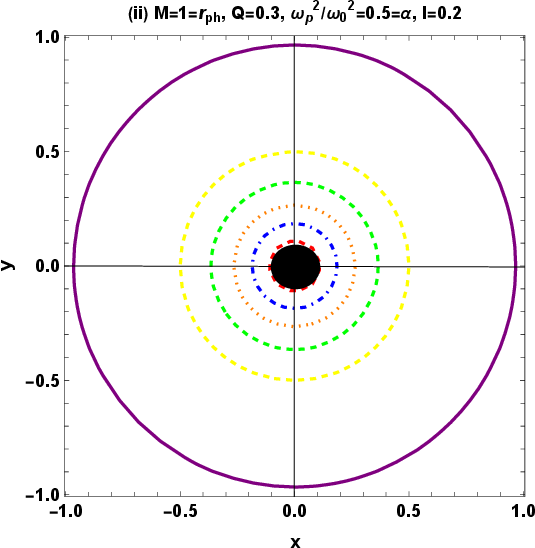}
\caption{The shadow of a hairy BH for different choices of radius $r_{ph}$ and $r_0$ and fixed values of
$M=1,~\alpha=0.5=\omega^2_{p}/\omega^2_{0}$, $l=0.2$ and Q=0.3.}\label{C1}
\end{figure}

The contour plots of shadow for various photon radius selections, $r_{ph}=0.5$ (purple), $0.6$ (yellow), $0.7$ (green), $0.8$ (orange), $0.9$ (blue), and $1$ (red), are shown in the left plot of Fig. {\ref{C1}}. It can be seen that as the photon radius increases, the shadow radius's size decreases.

The contour plots of shadow for various changes of radius $r_{0}=1$ (purple), $1.1$ (yellow), $1.2$ (green), $1.3$ (orange), $1.4$ (blue), and $1.5$ (red) are displayed in the right plot of Fig. {\ref{C1}}. It is evident that as the observer radius $r_{0}$ rises, so does the shadow radius's size.

\begin{table}[h]
  \centering
  \begin{tabular}{|c|c|c|c|c|c|c|}
    \hline
   $~~~~~r_{ph}~~~~~$ & ~~~~~0.5~~~~~ & 0.6 & 0.7 & 0.8 & 0.9 & 1 \\
    \hline
    $\beta$ & 0.159244 & 0.249447 & 0.376102 & 0.558127 & 0.831966 & 1.27735 \\
    \hline
  \end{tabular}
  \caption{Photon radius $r_{ph}$ and the shadow radius $\beta$ for fixed $M=1=r_0,~\alpha=0.5=\omega^2_{p}/\omega^2_{0}$, $l=0.2$ and Q=0.3.}
  \label{tab1}
\end{table}

\begin{table}[h]
  \centering
  \begin{tabular}{|c|c|c|c|c|c|c|}
    \hline
   $~~~~~r_{0}~~~~~$ & ~~~~~1~~~~~ & 1.1 & 1.2 & 1.3 & 1.4 & 1.5 \\
    \hline
    $\beta$ & 1.27735 & 0.507464 & 0.252109 & 0.133872 & 0.0707964 & 0.0345891 \\
    \hline
  \end{tabular}
  \caption{Observer radius $r_{0}$ and the shadow radius $\beta$ for fixed $M=1=r_{ph},~\alpha=0.5=\omega^2_{p}/\omega^2_{0}$, $l=0.2$ and Q=0.3.}
  \label{tab2}
\end{table}

\subsection{Emission Energy}

It is stated that the quantum fluctuations within hairy BH cause a significant amount of particles to form and be annihilated close to the horizon. Further, the positive-energy particles can escape the BH through tunneling within the space where Hawking radiation happens, which eventually causes the hairy BH to evaporate. We want to investigate the energy emission rate related to this case. For a distant observer, the high energy absorption cross-section approaches the hairy BH shadow. The hairy BH absorbed region oscillates at very high energies to a limiting constant value ($\sigma_{lim}$). It appears that the value of the limiting constant is correlated with the photon sphere's radius \cite{D60, D61} as
\begin{equation}
\sigma_{lim}\simeq \pi \beta^2,\label{E1}
\end{equation}
where $\beta$ denotes the hairy BH shadow radius and which gives the BH energy emission rate statement as
\begin{equation}
E_{\omega_{p},t}= \frac{2\pi\omega^{3}_{p}\sigma_{lim}}{e^{\frac{\omega_{p}}{T_{H}}}-1},\label{E2}
\end{equation}
with $\frac{d^{2}E(\omega_{p}, t)}{d\omega_{p} dt}=E_{\omega_{p},t}$, $T_{H}=\frac{\kappa}{2\pi}$, $\kappa$ and $\omega_{p}$ indicate the photon emission energy, the Hawking temperature, the hairy BH surface gravity and photon frequency, respectively. We may also formulate a new emission energy expression by merging Eqs. (\ref{E1}) and (\ref{E2}) by
\begin{equation}
E_{\omega_{p},t}= \frac{2\pi^{3}\omega^{3}_{p}\beta^2}{e^{\frac{\omega_{p}}{T_{H}}}-1}.\label{E3}
\end{equation}
For emission energy, we can determine the corresponding Hawking temperature \cite{RA1} for hairy BHs at the outer horizon ($r_{+}$) as
\begin{equation}
T_{H}= \frac{1}{4\pi r^{3}_{+}}\left[(2+\alpha l)Mr_{+}-2Q^{2}\right],\label{T_{H}}
\end{equation}
The mass of the BH, BH charge, geometric deformation of hairy BH parameters, a charge of hairy BH, and the outer horizon all affect the Hawking temperature. We note that the obtained Hawking temperature is known as the Schwarzschild BH for $Q=0$, $\alpha=0$, and $r_{+}=2M$.

\begin{figure}[H]
\centering
\includegraphics[width=6cm,height=6cm]{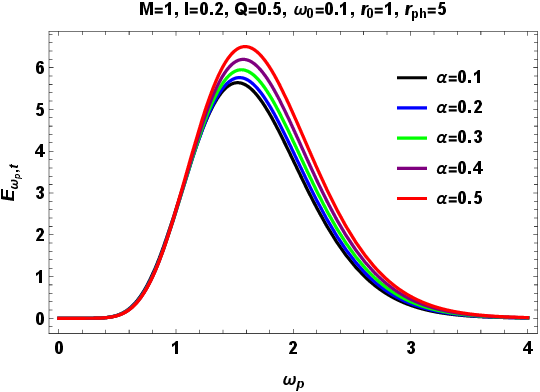}\includegraphics[width=6cm,height=6cm]{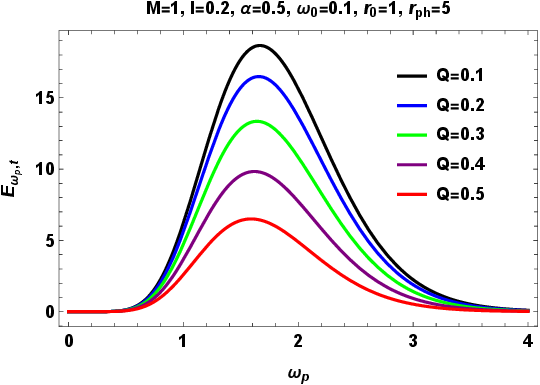}
\caption{Plots displaying the energy emission rate $E_{\omega_p,t}$ as a function of frequency for various values of the BH geometry parameter $\alpha$ (Left panel) and BH charge $Q$
(Right panel).}\label{e1}
\end{figure}

Fig. \ref{e1} shows the variation of energy emission with $\omega_p$ for fixed values of mass $M=1$, charge $Q=0.5$, observer frequency $\omega_0=0.1$, BH geometry parameter $l=0.2$, and radius $r_{ph}=5, r_0=1$ at different values of BH geometry parameter $\alpha$ (left panel), whereas different charge $Q$ and fixed $M=1$, charge $\alpha=0.5$, observer frequency $\omega_0=0.1$, BH geometry parameter $l=0.2$ and radius $r_{ph}=5, r_0=1$ (right panel). It is observed that with the increase in the values of $\alpha$, the peak of energy emission rate increases, while for the rise in BH charge $Q$, the peak of energy emission rate decreases, implying that the lower energy emission rate corresponds to the slow evaporation process in BH.

\section{Conclusion}

In this study, the deflection of light particles in curved hairy BH spacetime has been extensively investigated through the Gibbons-Werner strategy. A geometric expression and calculation of the deflection angle can be obtained by applying the GBT to regions formed in a two-dimensional manifold. Usually, an infinite region is selected for asymptotically flat hairy BH spacetimes. Although the singular behavior of the hairy BH metric causes the infinite region to be ill-defined for some asymptotically hairy BH spacetimes, several special regions have been developed to solve the ill-defined condition. We studied a generalized approach that greatly increases the option flexibility in the region where the GBT has been implemented. We applied the Gaussian curvature to compute the weak bending angle of hairy BH. We used the GBT to calculate the weak bending angle, which Gibbons-Werner initially studied. We studied the hairy spacetime theory and the optical geometry of BH in this way. From this point on, we have computed the bending angle obtained by the preceding order terms and used the GBT with a straight-line model. By integrating a domain outside the impact parameter, plasma terms and their effects on the bending angle of photons by hairy BH have also been examined. We analyzed from our findings that the mass of the BH $M$, BH charge $Q$, geometric deformation of hairy BH parameters $\alpha,~ l$, and impact parameter $b$ all affect the angle of light deflection that is obtained in a non-plasma medium. We noted that the first term in the obtained angle of light deflection is known as the Schwarzschild BH solution, and the others are caused by the BH's charged nature and the geometric deformation of hairy parameters. Furthermore, it should be observed that the second term positive sign suggests that this charged BH strong angle of light deflection is greater than that of the Schwarzschild BH, and the term negative sign suggests that this charged BH weak angle of light deflection. We also investigated the graphical effects of $Q$ and $\alpha$ on the angle of light deflection $\Psi$ to impact $b$. We observed that the charge ($Q$) has a minor effect on the angle of light deflection with a fixed deformation parameter ($\alpha$). The deformation parameter ($\alpha$) also has a minor effect on the angle of light deflection with a fixed charge. The graphs show that the deflection angle decreases exponentially with positive values before reaching an asymptotically flat form until it approximates the large value of $b$. We also observed that the deflection angle is greater than that of the Schwarzschild case in the plots for the non-plasma case. We also observed that for large values of charge $Q\geq1$, the angle is less than the Schwarzschild case. When we examined our results in a non-magnetic plasma medium, we concluded that the angle of light deflection depends on the impact parameter $b$, the geometry of BH $\alpha,~l$, charge $Q$, electron plasma frequency $\omega_e$, and photon frequency $\omega_{p}$ seen to infinity by the observer. If $\frac{\omega^{2}_{e}}{\omega^{2}_{p}} \rightarrow 0$, our results can be reduced to Eq. (\ref{a15}) in the non-plasma case.
We also analyzed our results in graphical form to check the effects of plasma medium on BH geometry. The graphs show that the light deflection angle reduces exponentially with positive values before reaching an asymptotically flat form until $b \rightarrow \infty$. It is also possible to see that the angle of light deflection grows with increasing deformation parameters ($\alpha$) and plasma frequency values. The light deflection angle in the Schwarzschild BH case is lower than that of hairy BH. We have also analyzed our analytical and graphical results in dark matter medium for deflection angle. We noted that weak and low deflection in the dark media is equivalent to that of the non-plasma and plasma mediums. The hairy BH instance has a greater deflection angle than the Schwarzschild BH example.

We presented the idea of a BH shadow and its significant use in astrophysics and BH physics. The critical sufficient condition for the separability of the Hamilton equation was obtained by applying ray-tracing of bent light rays in a non-magnetic plasma in hairy spacetime. By applying the ray-tracing approach to get a hairy BH shadow as a scenario, we checked the photon sphere radius and the observer radius effect on the shadow. We stated that the photon sphere was unstable in terms of photon orbits. The shadow becomes notably more significant with higher plasma frequency ratio values. The propagation in shadow $\beta$ as $r_{ph}$ grows may be explained by the direct relation between shadow and photon radius. Moreover, the shadow continuously grows for various values of deformation parameter $\alpha$, resulting in propagation in the BH shadow radius. The shadow is notably more enormous for larger values of the deformation parameter $\alpha$. Furthermore, from the contour plots in celestial coordinates $x$ and $y$, we concluded that for various photon radius selections and fixed choices of other parameters, the shadow radius's size decreases when the photon radius increases. From contour plots of shadow for multiple changes of radius $r_{0}$ and fixed choices of remaining parameters, when the observer radius $r_{0}$ rises, the shadow radius's size also rises. Moreover, we have studied the energy emission rate for the hairy BH. The dependence of the energy emission rate on the frequency
is analyzed in Fig. \ref{e1}. It is noted that for large values of $\alpha$ and small values of $Q$, the emission rate is more extensive. Thus, a significant quantity of energy is released with high $\alpha$ and low charge $Q$.

In the future, the deflection angles and shadow results gained can be applied to magnetized plasma surrounded by hairy BH in GR using the Gibbon-Werner and ray tracing techniques.
\section*{Acknowledgement}

The paper was funded by the National Natural Science Foundation of China 11975145.

\section*{Data Availibility Statement} The data that support the findings of this study are available from the corresponding
author upon reasonable request.

\section*{Declarations}
{\bf Conflict of interest} The authors declare no conflict of interest.

\end{document}